\documentclass[apj]{emulateapj}
\usepackage{bm}
\usepackage{graphicx}
\usepackage{epsf}
\usepackage{graphics}
\usepackage{amsmath}

\def\mpc{\,h^{-1}{\rm Mpc}}

\def\msun{\,h^{-1}{\rm M}_\odot}
\def\mstar{\,h^{-2}{\rm M}_\odot}

\def\cluster{{\tt cluster}}
\def\sheet{{\tt sheet}}
\def\filament{{\tt filament}}
\def\void{{\tt void}}

\newcommand{\kmsmpc}{\>{\rm km}\,{\rm s}^{-1}\,{\rm Mpc}^{-1}}

\usepackage{color}

\makeatletter

\newcommand{\Rmnum}[1]{\expandafter\@slowromancap\romannumeral #1@}
\makeatother

\shorttitle{size distributions of galaxies}
\shortauthors{Zhang et al.}

\begin{document}

\title{Size distribution of galaxies in SDSS DR7: weak dependence 
on halo environment}

\author{Youcai Zhang\altaffilmark{1}, Xiaohu Yang\altaffilmark{2,3}}

\altaffiltext{1}{Key Laboratory for Research in Galaxies and Cosmology,
  Shanghai Astronomical Observatory; Nandan Road 80, Shanghai 200030,
  China; E-mail: yczhang@shao.ac.cn}

\altaffiltext{2}{Department of Astronomy, Shanghai Jiao
Tong University, Shanghai 200240, China; Email: xyang@sjtu.edu.cn}

\altaffiltext{3}{IFSA Collaborative Innovation Center, Shanghai Jiao Tong
University, Shanghai 200240, China}

\begin{abstract}
  Using a sample of galaxies selected from the Sloan Digital Sky
  Survey Data Release 7 (SDSS DR7) and a catalog of bulge-disk
  decompositions, we study how the size distribution of galaxies
  depends on the intrinsic properties of galaxies, such as
  concentration, morphology, specific star formation rate (sSFR),
  and bulge fraction, and on the large-scale
  environments in the context of central/satellite decomposition, halo
  environment, the cosmic web: \cluster, \filament, \sheet ~and \void,
  as well as galaxy number density.  We find that there is a strong
  dependence of the luminosity- or mass-size relation on the galaxy
  concentration, morphology, sSFR, and bulge fraction.  
  Compared with late-type (spiral) galaxies, there is a
  clear trend of smaller sizes and steeper slope for early-type
  (elliptical) galaxies. Similarly, galaxies with high bulge fraction
  have smaller sizes and steeper slope than those with low bulge
  fraction. Fitting formula of the average luminosity- and mass-size
  relations are provided for galaxies of these different intrinsic
  properties.  Examining galaxies in terms of their large scale
  environments, we find that the mass-size relation has some weak
  dependence on the halo mass and central/satellite segregation for
  galaxies within mass range $9.0\le \log M_{\ast} \le 10.5$, where
  satellites or galaxies in more massive halos have slightly smaller
  sizes than their counterparts.  While the cosmic web and local
  number density dependence of the mass-size relation is almost
  negligible.
\end{abstract}

\section{Introduction}\label{sec_intro}

In the current paradigm of galaxy formation, galaxies are considered
to form from the accretion of gas in the gravitational potential well
provided by dark matter \citep{White1978}. According to the tidal
torque theory, dark matter and gas acquire their angular momenta by
the interaction between the inertia tensor and the local tidal field
\citep{Peebles1969, White1984}. In this scheme, the rotational discs
are assumed to form from the collapse of gas in the dark matter halos
\citep{Fall1980, Fall1983}. In the model for the formation of galactic
disc formulated by \citet{Mo1998}, the present-day discs assembled
recently at $z \le 1$. Discs in high redshift are small and dense, and
could merge together to form elliptical galaxies.

As one of the fundamental observational properties, the sizes of
galaxies are useful to calibrate the galaxy evolution models
\citep{Dejong2000, Kho2006, Truj2006, Truj2007, Buit2008, Hopkins2010,
 Roman2012, Capp2013, Fall2013, Java2017, Hill2017, Yild2017}. The sizes of
galaxies are well known to be correlated with their stellar mass (or
luminosity). Over the past few decades, a number of authors have
studied the luminosity- or mass-size relation of galaxies
\citep{Shen2003, Truj2004, Guo2009, Lange2015, Chan2016, Sweet2017,
  Furlong2017}. It's found that the relationship between size and mass
is strongly dependent on the morphology of the galaxy. At fixed
stellar mass, early-type (elliptical) galaxies have smaller sizes than
late-type (spiral) galaxies. In previous studies, the mass-size
relation has been fitted using single or double power law formula. For
early-type galaxies, \citet{Shen2003} used a single power law formula
to fit the mass-size relation \citep[see also][]{Wel2008,
  Bottrell2017}.  However, some studies showed that the most massive
part of the mass-size relation is curved in such a way that the
mass-size relation of the most massive galaxies have larger slope than
low-mass galaxies \citep{Desr2007, Hyde2009, Bernardi2011,
  Furlong2017}.  The steeper slope at high-mass end is likely caused
  by the larger sizes of central galaxies in clusters, i.e. cD galaxies,
  which have very extended luminosity profile so that have larger sizes 
  than the normal ellipticals at the same stellar mass.
In this case, the mass-size relation of early-type (elliptical)
galaxies can be well fitted by a double power law formula. For
late-type galaxies, most of studies used the double power law formula
to fit the mass-size relation \citep[e.g.,][]{Shen2003, Dutton2011},
whereas some authors used a single power law formula
\citep[e.g.,][]{Dutton2007, Bottrell2017}. In view of this, both
single and double power law functions are used by
\citet{Lange2015,Lange2016}.  They claimed that the single power law
formula is sufficient to describe the mass-size relation of late-type
galaxies, whereas the double power law is more robust than single
power law to fit early-type galaxies \citep[see Table $2$
of][]{Lange2016}.

More recently, a growing number of authors have endeavored to examine
the environmental dependence of the mass-size relation. Generally, the
environmental dependence of the luminosity- or mass-size relation is
controversial. On one hand, several studies have claimed that there is
no environmental dependence of the mass-size relation \citep{Rett2010,
  Maltby2010, Nair2010, Huertas2013, Kelkar2015, Sara2017}. Using $45$
massive ($M_{*}> 5 \times 10^{10} {\rm M}_{\odot}$) early-type
galaxies at $z \sim 1.2$, \citet{Rett2010} found that early-type
galaxies, both in clusters and in the field, follow similar mass-size
relation. Using a sample of $\sim 1200$ field and cluster galaxies,
\citet{Maltby2010} showed that there is no environmental dependence
for elliptical galaxies and for high-mass spiral galaxies
($M_{*}>10^{10} {\rm M}_{\odot}$). Using $12,150$ SDSS galaxies with
visual classification, \citet{Nair2010} showed that the slope of the
luminosity-size relation seems independent of environmental density.
Using a sample of $\sim 12000$ local early-type galaxies from SDSS
DR7, \citet{Huertas2013} claimed that galaxies in clusters have
similar sizes as the ones in the field. Using $\sim 1500$ galaxies at
$0.4 < z < 0.8$ from ESO Distant Cluster Survey, \citet{Kelkar2015}
found that there is no significant difference in the size
distributions of cluster and field galaxies. Compared a sample of $56$
elliptical galaxies in clusters at $z \sim 1.3$ with $\sim 430$ field
galaxies from GOODS, COSMOS and CANDELS, \citet{Sara2017} found that
there is no difference in the size at fixed mass of galaxies in
clusters and in the field.

On the other hand, there are a number of studies suggesting that the
sizes of galaxies are dependent on their environments. Some studies
claimed that early-type galaxies in clusters are larger than their
counterparts in the filed \citep{Papo2012, Bass2013, Lani2013,
  Strazz2013, Delaye2014, Yoon2017}. Using $< 100$ galaxies at
$z \sim 1.6$ from the CANDELS observation, \citet{Papo2012} and
\citet{Bass2013} claimed that quiescent galaxies in clusters are on
average larger compared with those in the field. Using $\sim 96,000$
galaxies from the UKIDSS Ultra Deep Survey, \citet{Lani2013} showed
that passive galaxies in high density environment are on average
significantly larger at $1<z<2$. Using $~12$ galaxies in a
X-ray-detected galaxy cluster at $z \sim 2$, \citet{Strazz2013} found
that passive early-type galaxies are larger by a factor of $\sim 2$ in
clusters than in the field. Using a sample of $\sim 400$ quiescent
early-type galaxies at $0.8<z<1.5$, \citet{Delaye2014} claimed that
the average sizes of galaxies in clusters are $30\%$-$40\%$ larger
than the ones in the field. In a recent study, \citet{Yoon2017} found
that early-type galaxies with mass larger than
$10^{11.2} {\rm M}_\odot$ in the high-density environments are as much
as $20\%$-$40\%$ larger than those in the low-density environments,
using $73,116$ early-type galaxies at $0.1 \le z < 0.15$ from SDSS
DR7. However, some studies indicated that galaxies in the field are
larger than the ones in clusters \citep{Raich2012, Pog2013, Cebri2014,
  Pran2017}.  Using a sample of $76$ early-type galaxies at
$z \sim 1.3$, \citet{Raich2012} found that galaxies in clusters are
smaller than field galaxies. Using a complete sample of galaxies at
$0.03 \leq z \leq 0.11$, \citet{Pog2013} indicated that galaxies in
the field are larger than in clusters. Based on galaxies from SDSS
DR7, \citet{Cebri2014} found that galaxies are larger in less-dense
regions than in high-density regions. Using about $700$ low-redshift
($z<0.063$) disc galaxies from SDSS DR7, \citet{Pran2017} found that
the sizes of galaxies are smaller by $\sim 15\%$ in clusters than in
the field.

In this study, using a sample of galaxies selected from SDSS DR7
\citep {Blanton2005} and a catalog of bulge-disk decompositions
\citep{Simard2011}, we investigate how the luminosity- or mass-size
relation depends on the intrinsic properties of galaxies, such as
concentration, morphology, specific star formation rate (sSFR), and 
bulge fraction, and on the large-scale environments
in the context of central/satellite decomposition, halo environment,
the cosmic web: \cluster, \filament, \sheet ~and \void, as well as
galaxy number density.  Here the cosmic web environments are
determined according to the eigenvalues of the tidal field,
constructed from the largest continuous region from SDSS DR7
\citep{Hahn2007a, Hahn2007b, Wang2012}.  While the central/satellite
separation and halo mass environment are calculated using galaxy
groups constructed by \citet{Yang2007}.

This paper is organized as follows. In Section~\ref{sec_data}, we
describe the observational data, including galaxies from NYU
Value-Added Galaxy Catalog and galaxies from bulge-disk decomposition
samples. In Section~\ref{sec_result}, we present how the luminosity-
or mass-size relation depends on galaxy morphology, bulge fraction,
and large-scale environments.  Finally, we summarize and discuss our
results in Section~\ref{sec_summary}. Unless stated otherwise, we
adopt a $\Lambda$CDM cosmology with parameters from \citet{Plank2016}:
$\Omega_{\rm m} = 0.308$, $\Omega_{\Lambda} = 0.692$,
$n_{\rm s}=0.968$, $h=H_0/(100 \kmsmpc) = 0.678$, and
$\sigma_8 = 0.815$.

\section{Observational Data}\label{sec_data}

In this section we describe the observational data we have used to
investigate the luminosity- or mass-size relations of
galaxies. Galaxies used in this paper come from the SDSS
\citep{York2000}, which has been one of the most successful surveys in
the history of astronomy. The SDSS has provided the most detailed
three-dimensional maps of the Universe, with deep multi-band images
and spectra for more than three million astronomical objects.

\subsection{NYU Value-Added Galaxy Catalog}

The galaxy sample used here is from the New York University
Value-Added Galaxy Catalog\footnote{http://sdss.physics.nyu.edu/vagc/}
\citep[NYU-VAGC;][]{Blanton2005}, which is based on the multi-band
imaging and spectroscopic survey SDSS DR7 \citep{Aba2009}. From the
NYU-VAGC, we collect a total of $639,359$ galaxies with redshifts in
the range $0.01\leq z \leq 0.2$ and with redshift completeness
${\cal C}_z > 0.7$.  For each galaxy, the $r$-band absolute magnitude
$M_{r}$ was computed, which was $K$-corrected and evolution corrected
to $z=0.1$ using the method described by \citet{Blanton2003} and
\citet{Blanton2007}.  In addition to the $r$-band absolute magnitudes,
we also use the stellar masses of galaxies.  Here the stellar masses 
and SFR of galaxies are obtained from the public catalog provided by
\citet{Chang2015}, in which we only use a total of $633,205$ galaxies
that have reliable aperture corrections (FLAG $=1$).

In the NYU-VAGC catalog, the Petrosian half-light radii $R_{50}$ and
$R_{90}$ are the radii enclosing $50\%$ and $90\%$ of the Petrosian
flux, respectively.  The Petrosian flux $F_{\rm P}$ in any band is
defined as the flux within a certain number $N_{\rm P}$
($N_{\rm P}=2.0$ in SDSS) of the Petrosian radius $r_{\rm P}$,
\begin{equation}\label{petro_flux}
F_{\rm P} = \int_0^{N_{\rm P} r_{\rm P}} 2\pi r I(r) {\rm d}r,
\end{equation}
where $I(r)$ is the azimuthally averaged surface brightness profile,
and $r_{\rm P}$ is defined as the radius at which the Petrosian ratio
$R_{\rm P}$ equals some specified value ($R_{\rm P}=0.2$ in SDSS).
The Petrosian ratio
$R_{\rm P}$ is defined as the ratio of the local surface brightness in
an annulus to the mean surface brightness within the radius
$r_{\rm P}$, which can be expressed by \citep{Blanton2001, Yasuda2001},
\begin{equation}\label{petro_radius}
R_{\rm P}= \frac {\int_{0.8r_{\rm P}}^{1.25r_{\rm P}} 2\pi r I(r)
{\rm d}r /[\pi(1.25^2-0.8^2)r_{\rm P}^2] } 
{\int_{0}^{r_{\rm P}} 2\pi r I(r) {\rm d}r /(\pi r_{\rm P}^2) } .
\end{equation}
Given the radii $R_{50}$ and $R_{90}$, the concentration index of the galaxy is
defined as $c=R_{90}/R_{50}$, which is correlated with galaxy morphological
type. As shown in \citep{Shen2003}, the SDSS galaxies may suffer small
fraction of incompleteness on sizes of galaxies, due to the very compact 
galaxies or very low surface brightness galaxies. In order to consider the
effect of the incompleteness on the luminosity-size relation, we have put
two reference lines (black dashed lines) on the left panel of 
Figure~\ref{fig:r50_c}, one is the $\mu_{\rm max} = 23.0 ~{\rm mag} ~{\rm arcsec}^{-2}$,
which corresponds to the very low surface brightness galaxies, and the
other is $\mu_{\rm min} = 18.5 ~{\rm mag} ~{\rm arcsec}^{-2}$, which may 
correspond a galaxy with $R_{50} = 2.0 ~{\rm arcsec}$ and the apparent magnitude
$m_r = 15$. As shown on the left panel of Figure~\ref{fig:r50_c}, most of galaxies
in our samples inside the two reference lines, which means that the incompleteness
have almost no effect on the luminosity- or mass-size relations, however, the
faint late-type galaxies ($c<2.85$) of $M_r>-18.0$ could be slightly biased by the 
incompleteness. Besides, we also adopt the morphological 
classifications of galaxies
from the Galaxy Zoo 2 Catalog \citep[GZ2;][]{Will2013}, which provides the most 
common classification for the galaxy with $gz2\_class$ strings. Elliptical
and spiral galaxies have $gz2\_class$ strings beginning with `E' and `S',
respectively. This results in $107,230$ elliptical galaxies and $134,024$ 
spiral galaxies cross-identified in our NYU-VAGC sample.

We also make use of the galaxy sample with bulge-disk decompositions
in the $r$-band for $1.12$ million galaxies from SDSS DR7
\citep{Simard2011}.  In their model, the galaxy image is fitted by the
sum of a pure exponential disk and a de Vaucouleurs bulge (S{\'e}rsic
index $n_b$). For comparison, \citet{Simard2011} used three different
fitting models: an $n_b=4$ bulge-disk model, a free-$n_b$ bulge-disk
model, and a pure S{\'e}rsic model. In the following analysis, we
adopt the galaxy structure parameters from their canonical $n_b=4$
bulge-disk fitting model using the GIM2D software package. In this
paper, the galaxy parameters we used are the $r$-band galaxy circular
half-light radius $R_{\rm chl}$, and the bulge-to-total ratio $B/T$ 
(see Table~$1$ in \citet{Simard2011}). From \citet{Simard2011}'s 
data base, $586,938$ (about $91.8\%$) galaxies can be cross 
identified within a total of $639,359$ galaxies in our SDSS DR7 
galaxy catalog.

\subsection{The large scale environments}

\begin{figure} \center \includegraphics[width=0.45\textwidth]{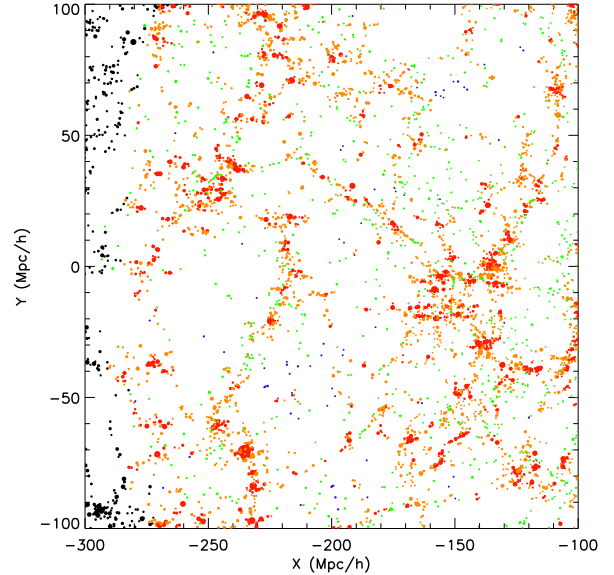}
  \caption{Spatial distribution of galaxies in different environments
    in a slice of thickness $20\mpc$ from SDSS DR7. The symbol sizes
    of the galaxies are proportional to their luminosities. The
    galaxies in four different environments are indicated by different
    colors: \cluster~(red), \filament~(orange), \sheet~(green) and
    \void~(blue). Black dots are galaxies outside the limited
    continuous volume, and are therefore not included in the
    analysis.}
\label{fig:slice}
\end{figure}

Based on the adaptive halo-based group finder developed by
\citet{Yang2005, Yang2007}, we construct a sample of $472,416$ groups,
among which $404,246$ groups are single galaxies. Using the complete
galaxy groups with masses $M_{\rm h} \geq 10^{12} \msun$,
\citet{Wang2012} constructed the tidal field $T_{i,j}$ with a
smoothing scale $R_s=2.1 \mpc$. According to the positive number of
the eigenvalues of local tidal tensor, the group's environment is
classified into one of four cosmic web types: \cluster, \filament,
\sheet, \void ~\citep{Hahn2007a, Hahn2007b, FR2009, Zhang2009,
  Zhang2013, Zhang2015}. If all of three eigenvalues at the position
of the group are positive, the group is classified into \cluster,
while the case of two, one or zero positive number of the eigenvalues
corresponds to \filament, \sheet ~or \void, respectively. In order to
ensure sample completeness, the galaxy's environment classification is
limited in the largest continuous region in the Northern Galactic Cap
of the SDSS DR7, which results in $117,667$ galaxies located in
\cluster, $212,075$ galaxies in \filament, $61,666$ galaxies in
\sheet, and $4,496$ in \void. Figure~\ref{fig:slice} shows the spatial
distribution of galaxies in a slice $200\mpc \times 200\mpc$ slice of
thickness $20\mpc$, in which galaxies in different environments are
indicated by different colors: \cluster ~(red), \filament ~(orange),
\sheet ~(green) and \void ~(blue).

In addition to the halo environment and cosmic web environment, we
also calculate the local (surface) number density of galaxies. The
surface number density of each galaxy is calculated by counting nearby
galaxies in a volume-limited sample: within the redshift range
$0.01<z<0.12$, and with the magnitude $M_r<-21.0$.  For each galaxy in
the redshift range $0.01<z<0.12$, we calculate the surface number
density by
\begin{equation}
\Sigma = n/\pi r^2,
\label{eqn:density}
\end{equation}
where $n$ is the galaxy counts ($M_r<-21.0$) in a cylinder of radius
$r$ and line-of-sight length $\Delta v$.  The unit of the surface
number density is ${\mpc}^{-2}$.

\begin{table}
\caption{Fitting parameters} 
\centering
\begin{tabular}{l c c c c}

\hline
\hline 
& \multicolumn{4}{c}{luminosity-size relation} \\
\cline{2-5}
Case & $\alpha$ & $\beta$ & $\gamma$ & $M_{r0}$ \\ 

\hline
Early type & 0.26 & 0.76 & 2.28 &-21.8 \\
Late type & 0.31 & 1.16 & 7.83  & -23.5 \\
Elliptical & 0.19 & 0.86 & 2.07 & -21.5 \\
Spiral &    0.26 & 2.33 & 8.53 & -24.5 \\
quiescent &  0.17 & 0.80 & 2.57 & -21.5 \\
star-forming & 0.37 & 0.42 & 11.95 & -23.9 \\
$B/T \geq 0.5$ & 0.28 & 0.84 & 1.86 & -21.4\\ 
$B/T < 0.5$ & 0.32 & 0.96 & 8.21 & -23.6\\ 
\hline
& \multicolumn{4}{c}{mass-size relation} \\
\cline{2-5}
Case & $\alpha$ & $\beta$ & $\gamma$ & $M_{0}$ \\ 
\hline
Early type & 0.11 & 0.60 & 1.75 & $1.35 \times 10^{10}\mstar$ \\
Late type & 0.22 & 1.24 & 8.83 & $4.49\times 10^{11}\mstar$ \\
Elliptical & 0.13 & 0.68 & 2.23 & $2.96 \times 10^{10}\mstar$  \\
Spiral & 0.16 & 5.41 & 8.96 & $1.93 \times 10^{12}\mstar$ \\
quiescent & -0.02 & 0.65 & 1.58 & $1.11 \times 10^{10}\mstar$ \\
star-forming & 0.23 & 0.41 & 10.72 & $4.90 \times 10^{11}\mstar$\\
$B/T \geq 0.5$ & 0.14 & 0.71 & 1.53 & $1.72 \times 10^{10}\mstar$\\ 
$B/T < 0.5$ & 0.18 & 0.78 & 6.34 & $1.57 \times 10^{11}\mstar$\\
 [1ex]
\hline

\end{tabular}
\label{table:parameters}
\end{table}

\section{Dependence on galaxy properties}\label{sec_result}

In this section, we investigate how the luminosity- or mass-size
relation depends on the intrinsic properties of galaxies, such as
concentration, morphology, specific star formation rate (sSFR), bulge
fraction and surface brightness.

\subsection{Dependence on Concentration}

\begin{figure*} \center \includegraphics[width=0.40\textwidth]{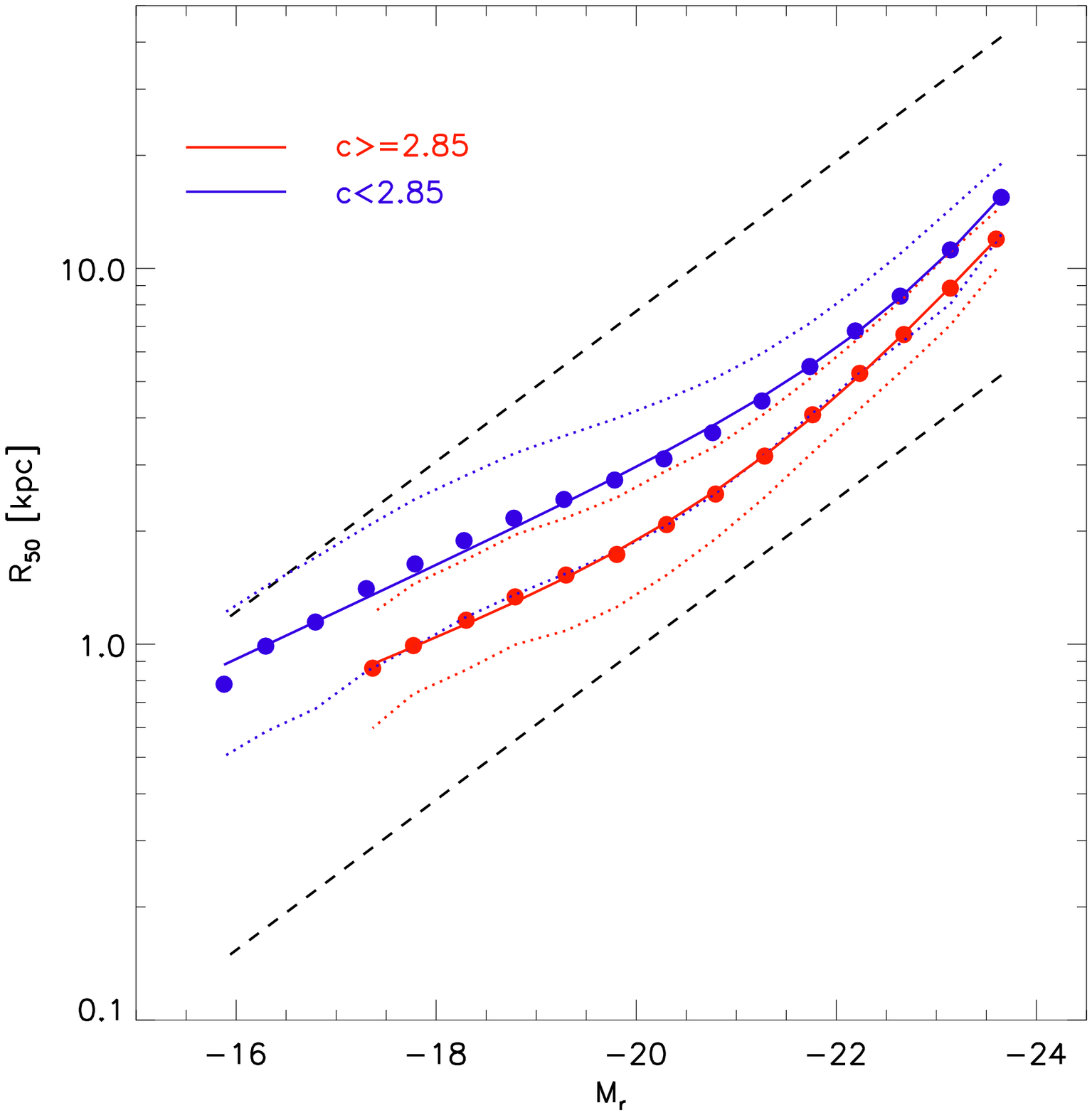}
\includegraphics[width=0.40\textwidth]{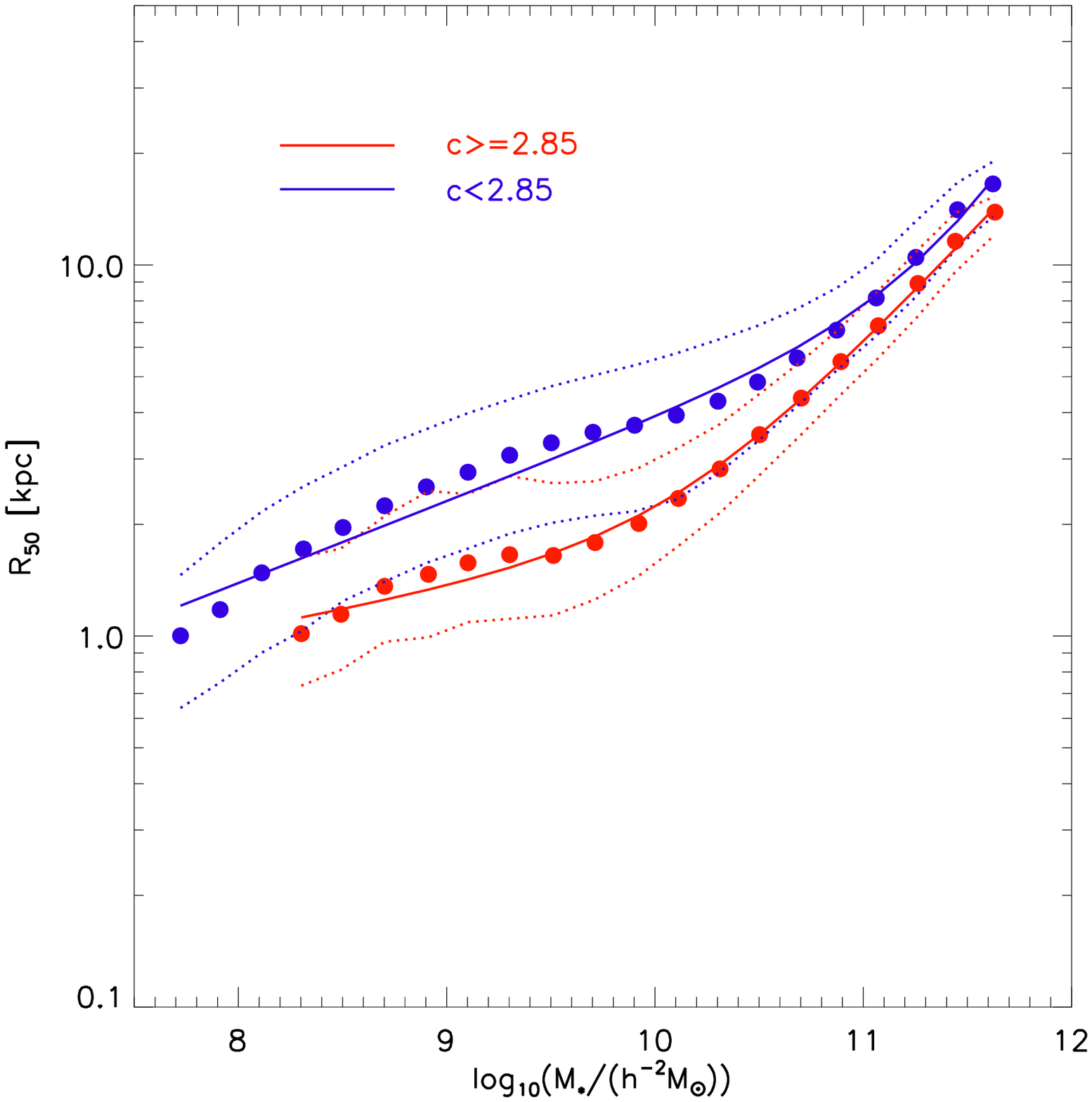}
\caption{Galaxy Petrosian half-light radius $R_{50}$ as a function of
  $r$-band absolute magnitude and stellar mass of galaxies in the
  NYU-VAGC samples. The red and blue solid points are median values of
  $R_{50}$ for galaxies with different concentration ($c \geq 2.85$ and
  $c<2.85$), and the corresponding dotted lines correspond to $16$ and
  $84$ percentiles. The solid lines show the fitting results using 
  Eqs.~\ref{eqn:r_m} and \ref{eqn:mass}, respectively. The black dashed
  lines correspond to the reference lines with $\mu_{\rm max} = 23.0 ~{\rm mag}
   ~{\rm arcsec}^{-2}$ and $\mu_{\rm min} = 18.5 ~{\rm mag} ~{\rm arcsec}^{-2}$. }
\label{fig:r50_c}
\end{figure*}

The concentration index, $c=R_{90}/R_{50}$, is found to be tightly
correlated with galaxy morphological type \citep{Shimasaku2001,
  Shen2003, Park2005, Deng2013, Deng2015}. \citet{Shimasaku2001}
demonstrated that the concentration index can be used to classify
galaxies into early and late types with their recommended choice of
$c=3.03$. \citet{Naka2003} separated galaxies into early and late
types with $c=2.857$. They claimed that this choice of the
concentration index $c=2.857$ can minimize the contamination of the
opposite morphological type. \citet{Shen2003} used $c=2.86$ to
separate galaxies into early and late types. Based on SDSS DR8,
\citet{Deng2013} claimed that the concentration index $c=2.85$ can be
used to construct a reasonably pure late-type galaxy sample, although
it's unfortunately not good enough to construct an early-type galaxy
sample. In this paper, we separate galaxies into early and late types
according to $c \geq 2.85$ and $c<2.85$. In the $r$-band absolute
magnitude range $-24.0 \leq M_{r} \leq -15.5$, the early-type
subsample ($c \geq 2.85$) contains $214,950$ galaxies, while the
late-type subsample ($c<2.85$) contains $424,363$ galaxies. Based on
these galaxy samples, we investigate the dependence of the
luminosity-size relation on the galaxy concentration index.

In the left panel of Figure~\ref{fig:r50_c}, we show the galaxy
Petrosian half-light radius $R_{50}$ as a function of $r$-band
absolute magnitude $M_{r}$ of galaxies in the NYU-VAGC samples from
SDSS DR7. The size distribution of galaxies at give luminosity 
(or stellar mass) can be well described by a log-normal distribution
with the median value of $R_{50}$. Therefore, in this paper we use 
the median values of $R_{50}$ to characterize the sizes of galaxies 
in each magnitude (or mass) bin. In Figure~\ref{fig:r50_c}, the red 
and blue solid dots denote the median values of $R_{50}$ for early-type 
($c \geq 2.85$) and late-type ($c<2.85$) galaxies, respectively, 
with the corresponding dotted lines representing $16$ and $84$ percentiles.

\begin{figure*}
\center
\includegraphics[width=0.40\textwidth]{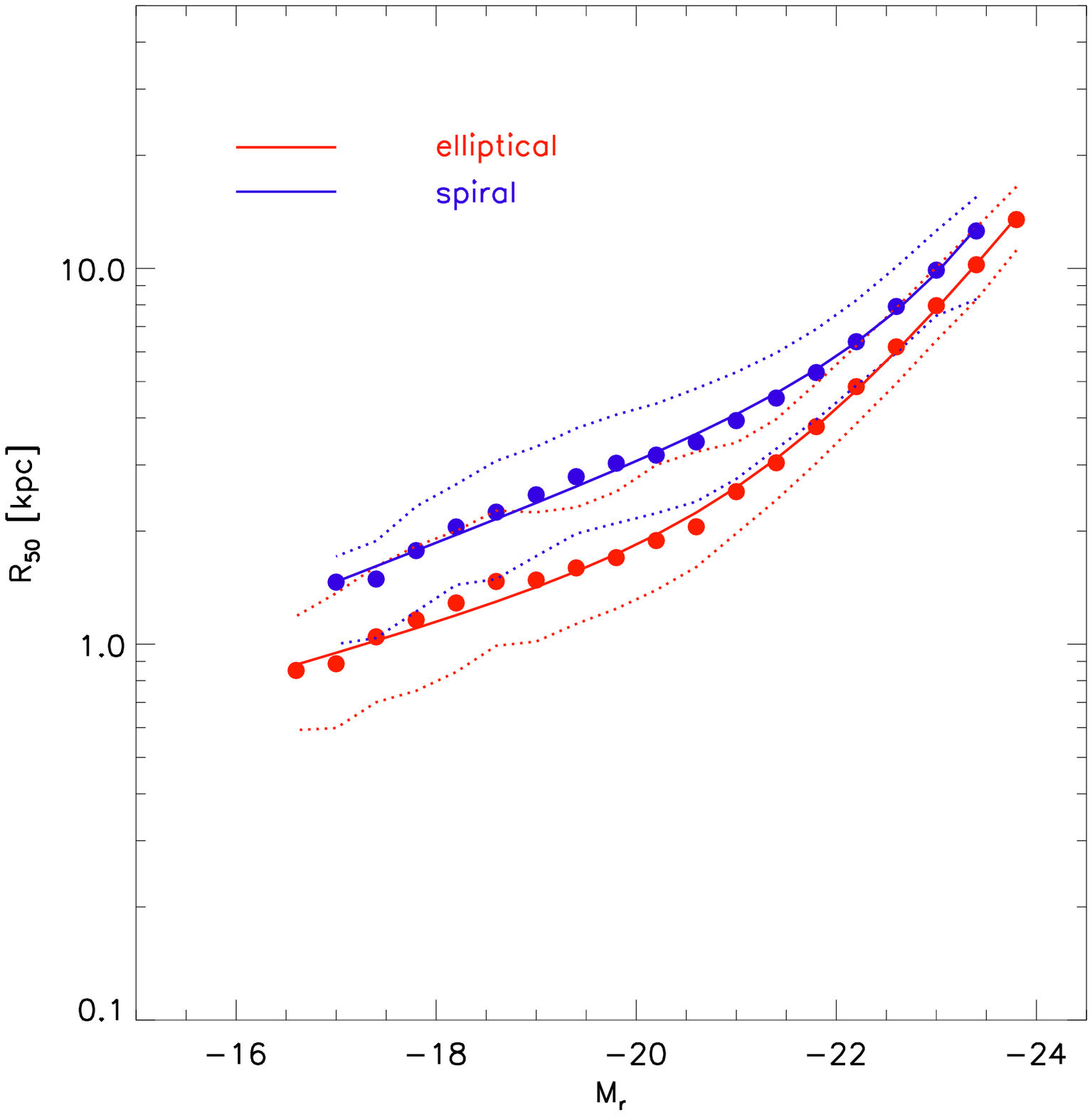}
\includegraphics[width=0.40\textwidth]{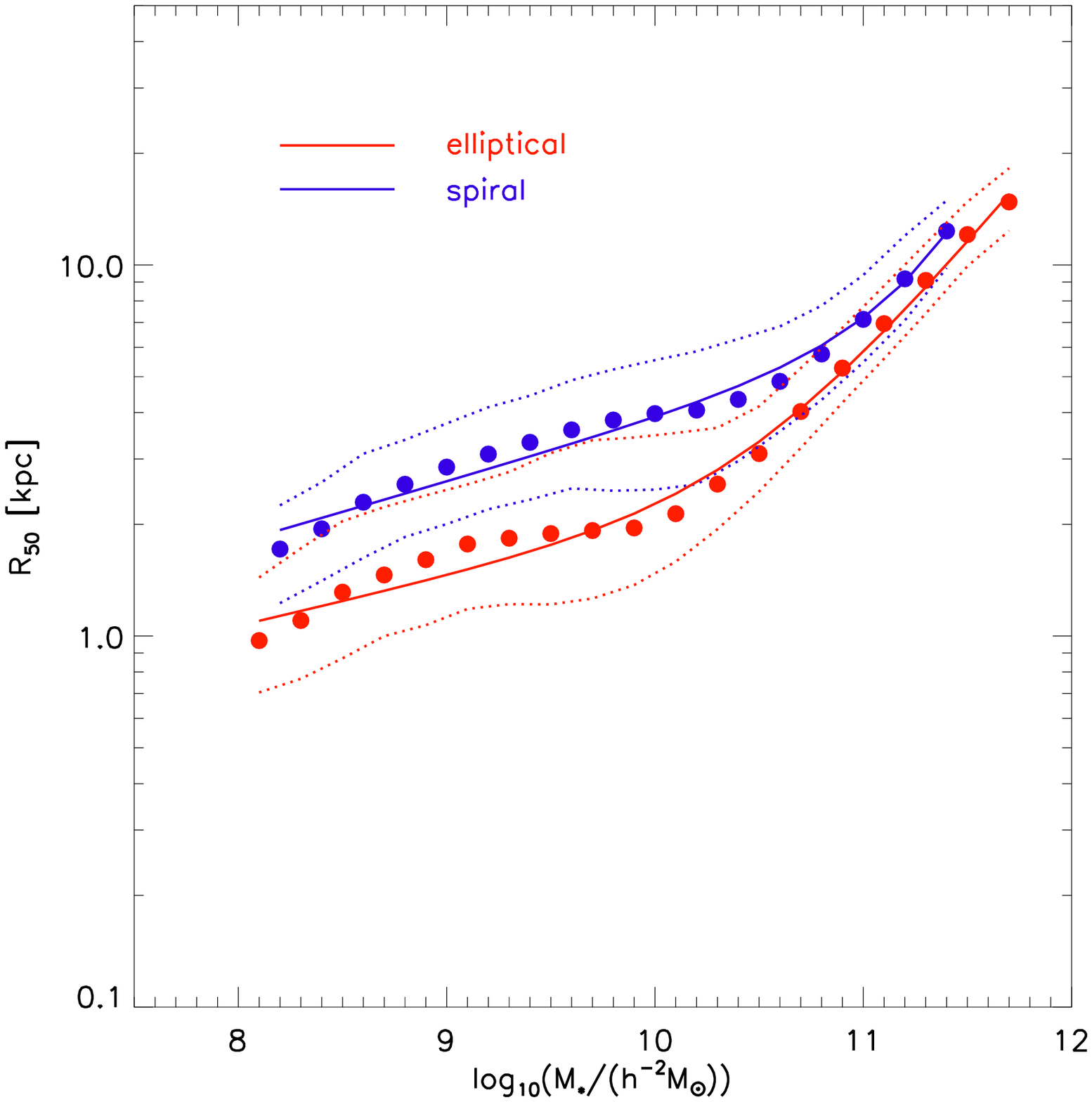}
\caption{Similar to Figure~\ref{fig:r50_c}, but for galaxy
  morphological classifications according to Galaxy Zoo 2 Catalog
  \citep{Will2013}. The red and blue data points are median values of
  $R_{50}$ for elliptical and spiral galaxies, respectively. }
\label{fig:r50_se} \end{figure*}

As shown in Figure~\ref{fig:r50_c}, the luminosity-size relations are
quite different for early-type and late-type galaxies. Generally,
early-type galaxies have smaller sizes than late-type galaxies, which
was also reported by \citet{Shen2003, Courteau2007, Bottrell2017}. To
quantify the relation between galaxy size $R_{50}$ and absolute
magnitude $M_{r}$, we employ the following simple formula
\begin{equation}\label{eqn:r_m}
R_{50} = \gamma L^\alpha (1+L)^{(\beta-\alpha)},
\end{equation}
where $L=10^{-0.4(M_r-M_{r0})}$ is proportional to the luminosity,
$\alpha$, $\beta$, $\gamma$, and $M_{r0}$ are four free fitting
parameters. The method of least-squares is used to estimate the
fitting parameters for the relation between the average of the galaxy
size $R_{50}$ and the $r$-band absolute magnitude $M_{r}$. The fitting
results are shown as the solid lines in Figure~\ref{fig:r50_c}.  For
$214,950$ early-type galaxies, we find that $\alpha=0.26$,
$\beta=0.76$, $\gamma=2.28$, and $M_{r0}=-21.8$ can provide a good fit
to the data, while for $424,363$ late-type galaxies, the fitting
parameters are $\alpha=0.31$, $\beta=1.16$, $\gamma=7.83$, and
$M_{r0}=-23.5$.  According to comparison between the data and fitting
results, we see that the model gives a very good description of the
luminosity-size relations.  In addition, one can see that the slopes
of the luminosity-size relations show a prominent dependence on galaxy
morphology. Generally, early-type galaxies have a deeper slope than
late-type galaxies. This result is in qualitative agreement with a
number of previous studies \citep[e.g.,][]{Shen2003, Courteau2007,
  Dutton2011}.

In addition to the absolute magnitudes, we also study the size
distribution of galaxies as a function of stellar mass. The mass-size
relations are shown in the right panel of Figure~\ref{fig:r50_c} as
solid dots. To quantify the mass dependence of $R_{50}$, we fit the
average mass-size relations by the following formula
\begin{equation}\label{eqn:mass}
R_{50} = \gamma \left( \frac {M_{*}}{M_0}\right)^{\alpha} 
\left( 1+ \frac {M_{*}} {M_0} \right)^{(\beta - \alpha)} ,
\end{equation}
where $M_{*}$ is the stellar mass of the galaxy, $\alpha$ is the slope
with stellar mass $M_{*} \ll M_{0}$, $\beta$ is the slope with stellar
mass $M_{*} \gg M_{0}$, and $M_{0}$ is the transition mass. Here,
$\alpha$, $\beta$, $\gamma$, and $M_{0}$ are all fitting
parameters. The least-squares method is used to fit these
parameters. The fitting results are shown as solid curves in
Figure~\ref{fig:r50_c}.  

For late-type galaxies (blue curve), the fitting parameters are
$\alpha=0.22$, $\beta=1.24$, $\gamma=8.83$, and
$M_0=4.49 \times 10^{11}\mstar$.  For early-type galaxies (red curve),
the fitting parameters are $\alpha=0.11$, $\beta=0.60$, $\gamma=1.75$,
and $M_0=1.35 \times 10^{10}\mstar$. The high-mass slope $\beta=0.60$
of early-type galaxies is almost consistent with the slope $0.57$
measured for quiescent high-mass galaxies
$M_* > 10^{10.7} {\rm M}_\odot$ from SDSS DR7 \citep{Newman2012}, and
with the slope $0.58$ measured from SDSS early-type galaxies with mass
$M_* > 10^{10.5} {\rm M}_\odot$ \citep{Cima2012}.

\begin{figure*} \center \includegraphics[width=0.40\textwidth]{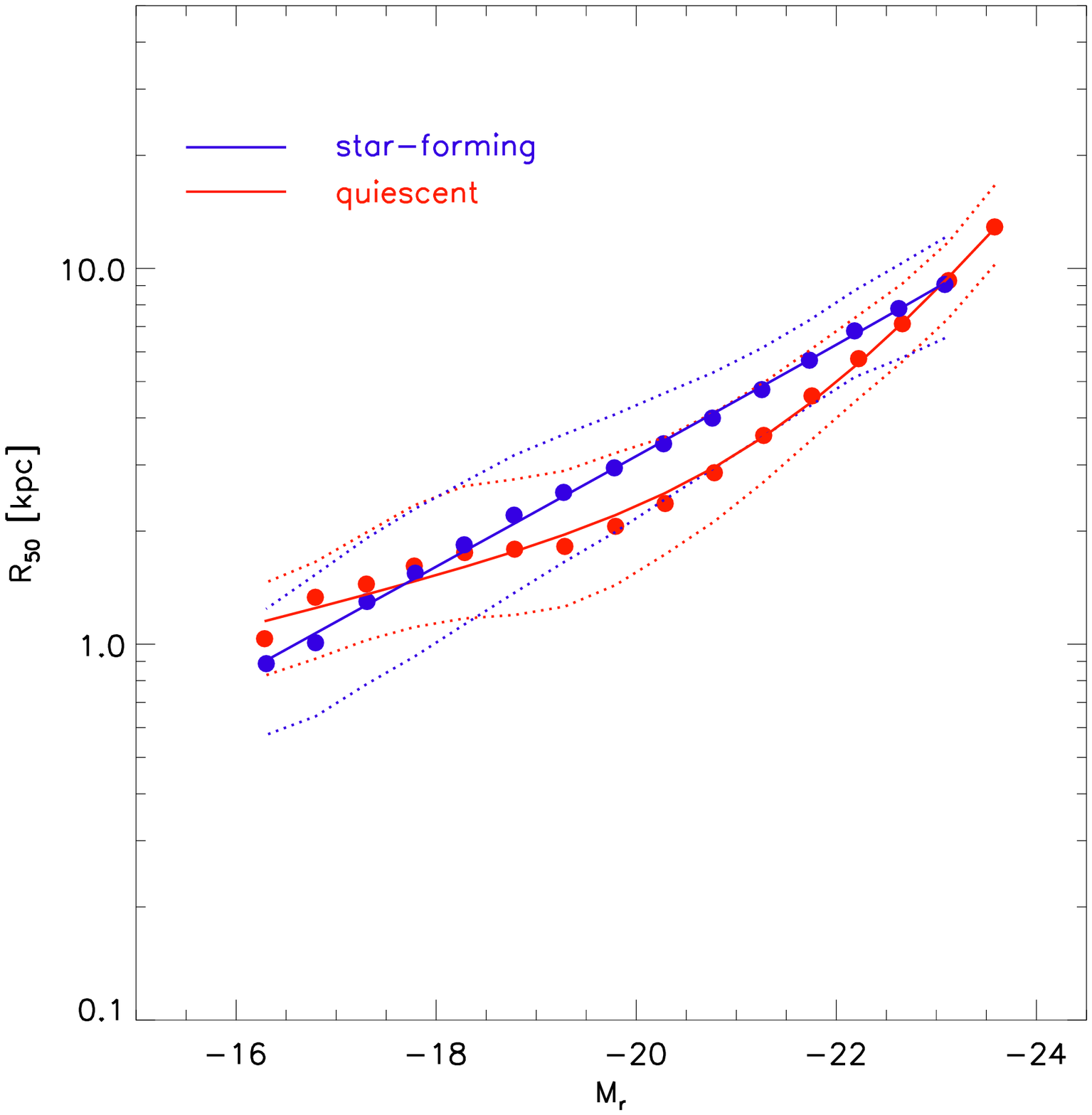}
\includegraphics[width=0.40\textwidth]{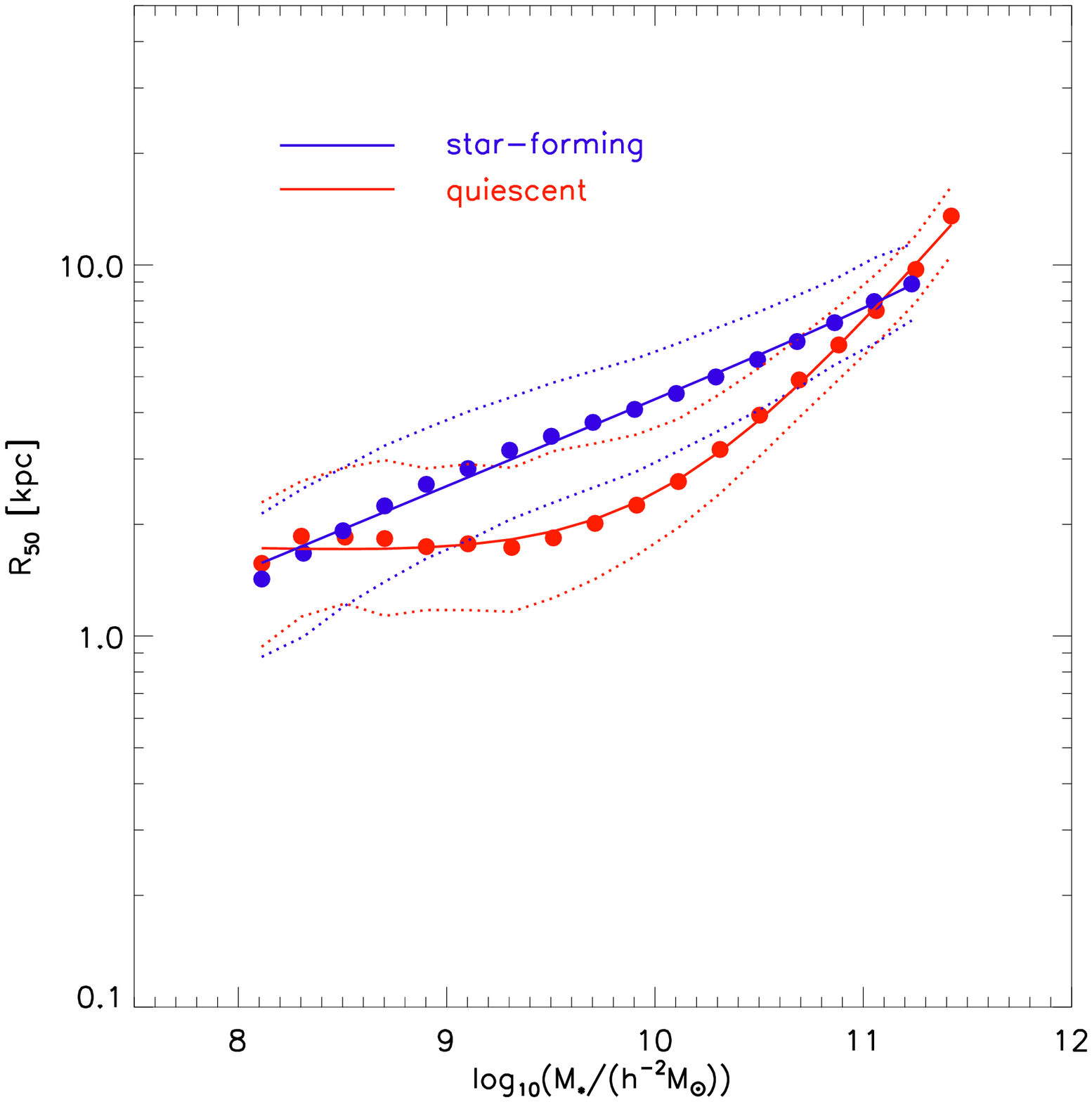}
\caption{Similar to Figure~\ref{fig:r50_c}, but for star-forming and
  quiescent galaxies. The blue and red data points are median values
  of $R_{50}$ for star-forming and quiescent galaxies,
  respectively. }
\label{fig:ssfr_mag}
\end{figure*}

\subsection{Dependence on morphology}

Next, we study the luminosity-size relation based on the morphological
classifications from Galaxy Zoo 2 Catalog \citep{Will2013}. The left
panel of Figure~\ref{fig:r50_se} shows the luminosity-size relations
for $107,230$ elliptical (red) galaxies and $134,024$ spiral (blue)
galaxies. As shown in Figure~\ref{fig:r50_se}, elliptical galaxies
have smaller size than spiral galaxies, which is similar to that based
on galaxy classifications according to the concentration $c$ criterion
in Figure~\ref{fig:r50_c}. Equation~\ref{eqn:r_m} is also used to fit
the $R_{50}-M_{r}$ relations, resulting in $\alpha=0.19$, $\beta=0.86$,
$\gamma=2.07$, $M_{r0}=-21.5$ for elliptical galaxies and
$\alpha=0.26$, $\beta=2.33$, $\gamma=8.53$, $M_{r0}=-24.5$ for spiral
galaxies.

The right panel of Figure~\ref{fig:r50_se} shows the size distribution
of galaxies as a function of stellar mass. Here again, we use the
least-squares method to fit these parameters in Eq. \ref{eqn:mass}.
The fitting results are shown as the solid curves in
Figure~\ref{fig:r50_se}. For spiral galaxies, the fitting parameters
are $\alpha=0.16$, $\beta=5.41$, $\gamma=8.96$, and
$M_0=1.93 \times 10^{12}\mstar$. For elliptical galaxies, the fitting
parameters are $\alpha=0.13$, $\beta=0.68$, $\gamma=2.23$, and
$M_0 = 2.96 \times 10^{10} \mstar$. Overall, the mass-size relations
for both spiral and elliptical galaxies follow the double power law
distribution quite well. For elliptical galaxies, the low-mass
galaxies ($M_* \ll M_0$) have $R_{50} \propto M_{*}^{0.13}$, and the
high-mass galaxies ($M_* \gg M_0$) have $R_{50} \propto
M_{*}^{0.68}$. \citet{Wel2014} also found a steep slope of
$0.75 \pm 0.06$ for massive ($M_*>2 \times 10^{10} {\rm M}_\odot$)
early-type galaxies at $z \sim 0.25$.

\subsection{Dependence on specific $\rm SFR$}

In addition to the concentration and morphology dependences, we also
probe the luminosity-size relations for star-forming and quiescent
galaxies.  Using the best-fitting formula obtained by
\citet[][Equation $4$]{Chang2015}, 
\begin{equation}\label{eqn:ssfr}
\log [{\rm sSFR}/({\rm yr}^{-1})] = -0.2 \log [ M_*/(h^{-2}{\rm M}_\odot)]
- 8.94\,,
\end{equation}
we separate galaxies in our sample into star-forming and quiescent
subsamples. Apart from this separation criteria, we also test the
separation criteria proposed by \citet{Brinchmann2004}, where galaxies
are separated into two distinct populations using a constant sSFR
value (e.g.  $\log {\rm sSFR} = -11.0$). We have checked our final
results using a constant sSFR value to distinguish star-forming and
quiescent galaxies, and found that the results are very similar to
those using Equation~\ref{eqn:ssfr}.

Figure~\ref{fig:ssfr_mag} show the luminosity-size relations and
mass-size relations for star-forming and quiescent galaxies in
NYU-VAGC samples. The fitting parameters using Equation~\ref{eqn:r_m}
and Equation~\ref{eqn:mass} are listed in
Table~\ref{table:parameters}\footnote{For clarity, in what follows we
  only list the best fitting parameters in Table
  \ref{table:parameters}.} .  Generally, the sizes of star-forming
galaxies are larger than those of quiescent galaxies. This result is
expected due to the fact that star-forming galaxies are less
concentrated, and quiescent galaxies are more concentrated
\citep{Brinchmann2004}.  Note that dwarf quiescent galaxies in the mass
range $10^{8.0} \mstar \le M_{*} < 10^{9} \mstar$ have larger sizes 
than galaxies with red solid points in the right panels of Figure~\ref{fig:r50_c},
~\ref{fig:r50_se}, and ~\ref{fig:rchl}. This might be because that dwarf
quiescent galaxies, which has exponential profile, are more likely separated
into late-type galaxies using concentration as indicator, while they are 
quiescent galaxies using sSFR as indicator. As can be seen, both 
star-forming and quiescent galaxies can be well fitted by the double power law
formula. Nevertheless, a single power law formula is also sufficient
to describe the luminosity- or mass-size relations of star-forming
galaxies. For example, the mass-size relation of star-forming galaxies
can be also well fitted by $\log R_{50} = a \log M_* + b$, where
$a=0.24$, and $b=-1.75$.

\subsection{Dependence on bulge fraction}

\begin{figure*}
\center
\includegraphics[width=0.40\textwidth]{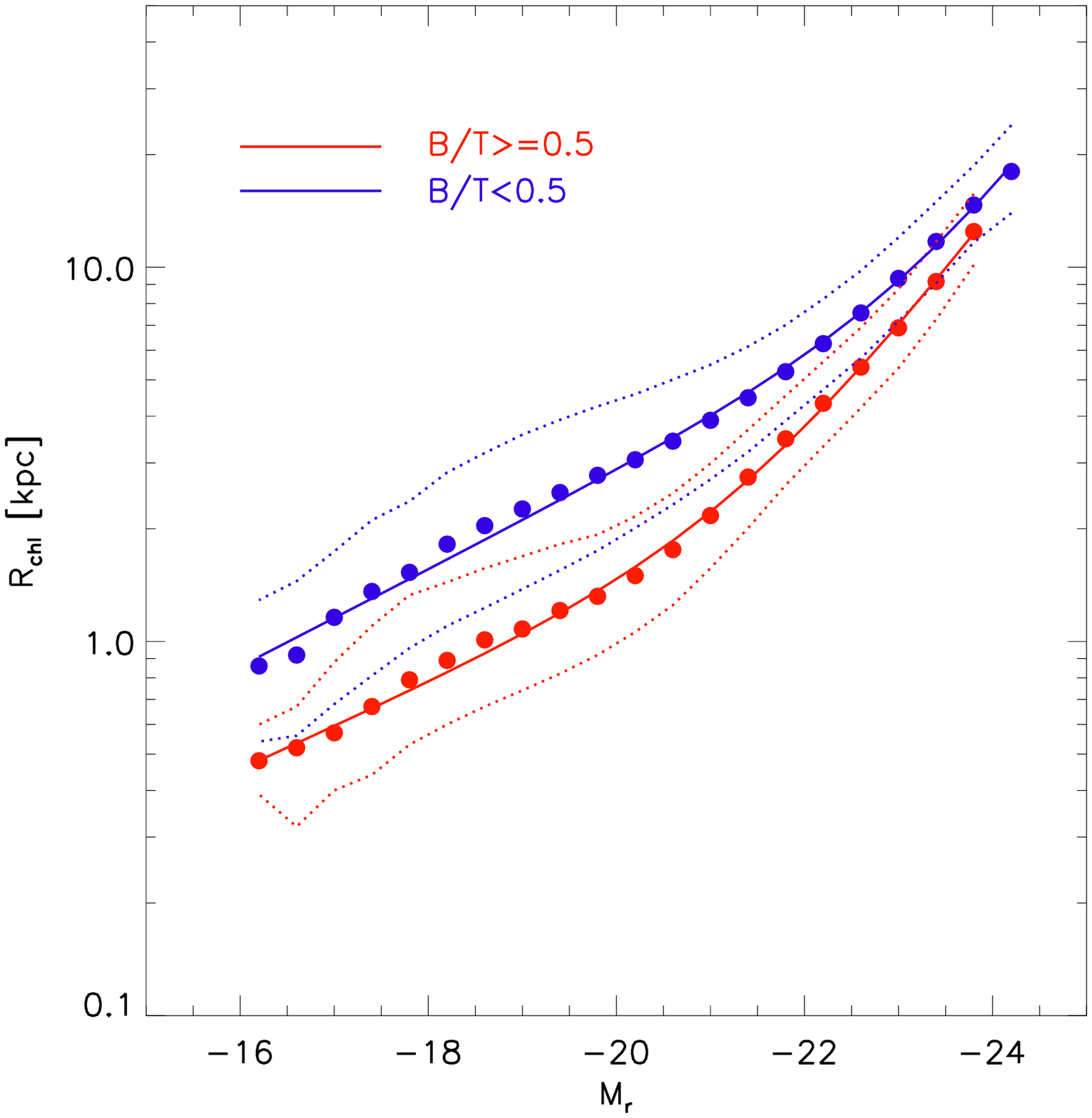}
\includegraphics[width=0.40\textwidth]{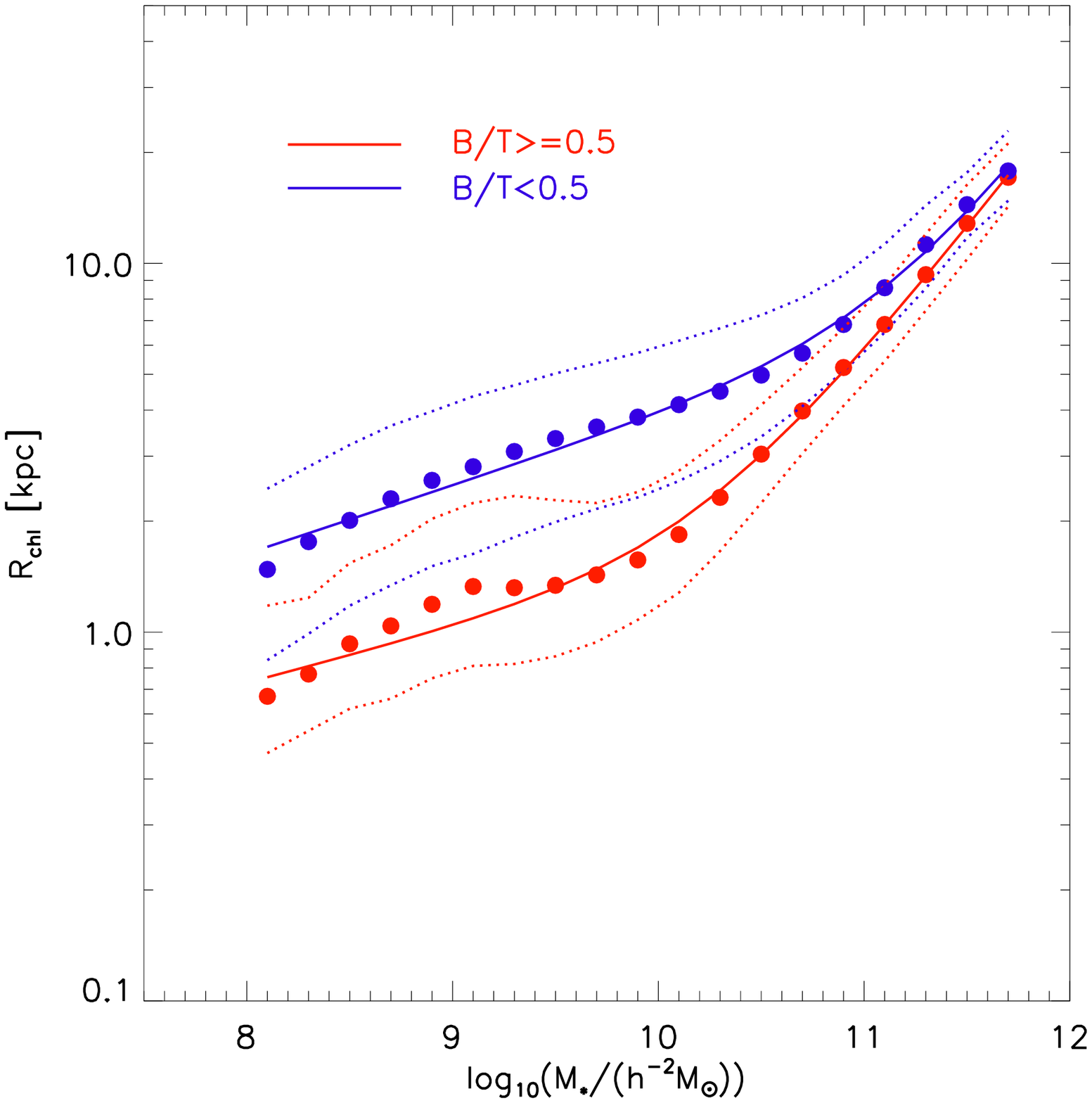}
\caption{Galaxy circular half-light radius $R_{\rm chl}$ as a function
  of $r$-band absolute magnitude and stellar mass of galaxies cross
  identified with \citet{Simard2011}'s data base. The blue and red data
  points are median values for galaxies with different $r$-band bulge
  fraction ($B/T<0.5$ and $B/T\geq0.5$). The solid
  lines show the fitting results using Eqs.~\ref{eqn:r_m} and \ref{eqn:mass},
  respectively. }
\label{fig:rchl}
\end{figure*}

In \citet{Simard2011}'s data base, the galaxy image is fitted by the
sum of a pure exponential disk and a de Vaucouleurs bulge. Using
$586,938$ galaxies cross identified from the \citet{Simard2011}'s data
base, we further examine the dependence of the luminosity-size
relation on the bugle fraction. In their canonical fitting model
(bulge S{\'e}rsic index $n_b=4$), \citet{Simard2011} used the version
$3.2$ of the software package GIM2D to calculate the galaxy structural
parameters, which are listed in Table 1 of \citet{Simard2011}. In this
subsection, the structural parameters we used are the $r$-band galaxy
circular half-light radius $R_{\rm chl}$ and the bulge fraction
$B/T$. In \citet{Simard2011}'s data base, the galaxy circular
half-light radius $R_{\rm chl}$ is calculated by integrating GIM2D
best-fit models with the summed bulge and disk profiles. Note that the
bugle fraction can be best linked to the galaxy morphology together by
the image smoothness \citep{Simard2002, Simard2009}. We have found
that the luminosity-size relation is strongly dependent on the
morphology. A similar dependence on the bulge fraction thus is
expected.

The left panel of Figure~\ref{fig:rchl} shows the galaxy circular
half-light radius as a function of $r$-band absolute magnitude of
$586,938$ galaxies cross identified in \citet{Simard2011}'s data
base. The relations of the galaxies with low ($B/T<0.5$) and high
($B/T \geq 0.5$) bulge-to-total ratios are denoted by blue and red
data points, respectively. The solid lines are the fitting results by
Equation~\ref{eqn:r_m}. The fitting parameters given by the
least-squares method are listed in Table~\ref{table:parameters}. As
one can see, the size differences between low and high bulge fraction
galaxies are somewhat similar to those of galaxies separated using
different concentrations in Figure~\ref{fig:r50_c}. At fixed absolute
magnitude, the galaxies of higher bulge fraction $B/T$ have smaller
sizes. Besides, the luminosity-size relation with high bulge fraction
($B/T\geq 0.5$) has a steeper slope, especially for brighter galaxies
($M_{r} \leq -20.5$).  The right panel of Figure~\ref{fig:rchl} shows
the galaxy circular half-light radius as a function of stellar
mass. The fitting results are shown as the solid lines. The fitting
parameters are also listed in Table~\ref{table:parameters}.  Similar
to the luminosity-size relation, galaxies with high bulge fraction
have steeper slope, especially for massive galaxies. This trend of
increasing slope and decreasing sizes of the galaxies with higher
bulge fraction agrees well with the results of \citet{Bottrell2017}
based on galaxy images from the Illustris simulation \citep{Vog2014}
and the SDSS.

\section{Dependence on Large-scale Environment}

Having modelled the luminosity- and mass-size relations for galaxies
of different intrinsic properties, we proceed to probe their
dependences on large scale environments.   For simplicity, we only
provides results based on the stellar mass of galaxies. Those of
luminosities are very similar.

\subsection{Dependence on halo environment}

\begin{figure}
\center
\includegraphics[width=0.45\textwidth]{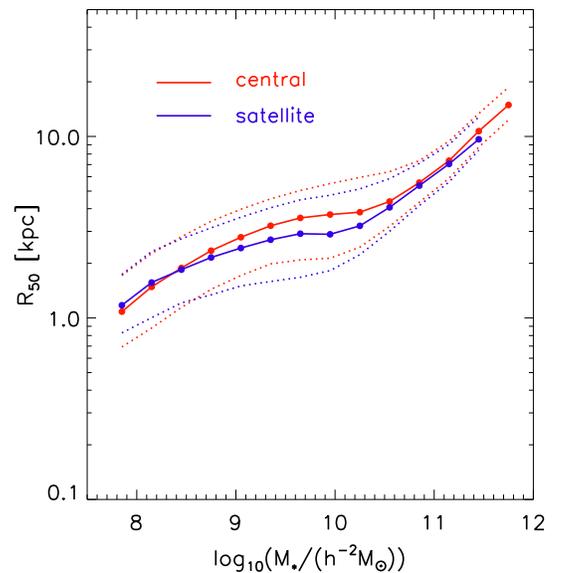}
\caption{Galaxy Petrosian half-light radius $R_{50}$ as a function of
  stellar mass for central and satellite galaxies.}
\label{fig:cs}
\end{figure}

\begin{figure}
\center
\includegraphics[width=0.45\textwidth]{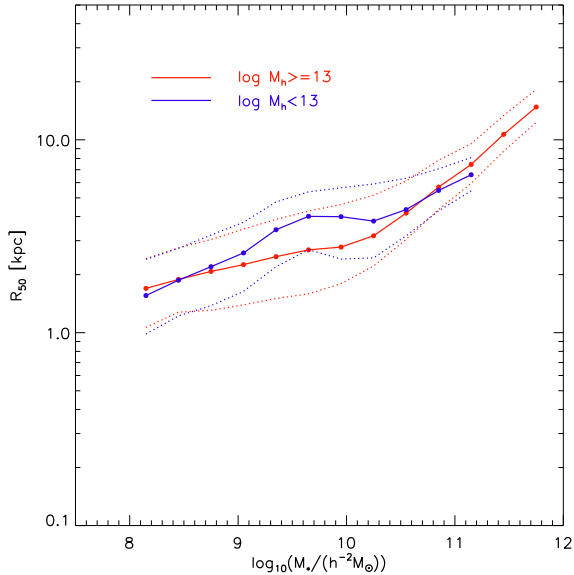}
\caption{Galaxy Petrosian half-light radius $R_{50}$ as a function of
  stellar mass for galaxies in halos with mass larger or smaller than
  $10^{13.0}\msun$. }
\label{fig:cs_hm}
\end{figure}

Note that all the results in this section are also investigated
using early- or late-type galaxies. The results are very similar to those 
using all the galaxies. Therefore, in this section, we only present
the results using all the galaxies for simplicity.

The first large scale environment we check is the halo environment.
Based on the group catalog provided by \citet{Yang2007} from SDSS DR7,
we separate the galaxies into centrals and satellites.
Figure~\ref{fig:cs} shows the mass-size relations for central and
satellite galaxies.  Here we find that the central galaxies have
slightly larger sizes in the mass range
$10^{9.0} \mstar \le M_{*} < 10^{10.5} \mstar$.  This dependence
disappears in galaxies beyond this mass range.  In addition, as we
have also tested by further separating the central and satellite
galaxies into early-type or late-type sub-subsamples, the
central/satellite dependence is quite similar. Using a sample of $911$
central galaxies from SDSS DR4, \citet{Guo2009} also found that there
are no size differences between early-type central and satellite
galaxies, especially in the mass range
$10^{10.5} \mstar \le M_{*} < 10^{11.25} \mstar$ (see the upper-right
panel of their Figure~$10$). \citet{Huertas2013} also claimed that
central and satellite galaxies follow similar mass-size relations,
based on $\sim 12000$ early-type galaxies with mass
$10^{10.5} \mstar \le M_{*}$ in SDSS DR7.

In addition to the central/satellite separation, we also probe the
mass-size relations for galaxies in high and low mass
groups/halos. Based on the halo masses estimated by \citet{Yang2007}
for galaxy groups in the SDSS DR7, we separate the galaxies into two
subsamples according to their halo mass. Figure~\ref{fig:cs_hm} shows
the mass-size relations of galaxies in halos with mass larger or
smaller than $10^{13.0}\msun$.  Here, we find galaxies in larger
halos have smaller sizes in the stellar mass range $10^{9.0} \mstar 
\le M_{*} < 10^{10.5} \mstar$.  This is expected because that galaxies
with $10^{9.0} \mstar \le M_{*} < 10^{10.5} \mstar$ in halos larger
that $10^{13.0}\msun$ are more likely belong to satellite galaxies. Therefore,
the results in Figure~\ref{fig:cs} and Figure~\ref{fig:cs_hm} are consistent.

\begin{figure*}
\center
\includegraphics[width=0.9\textwidth]{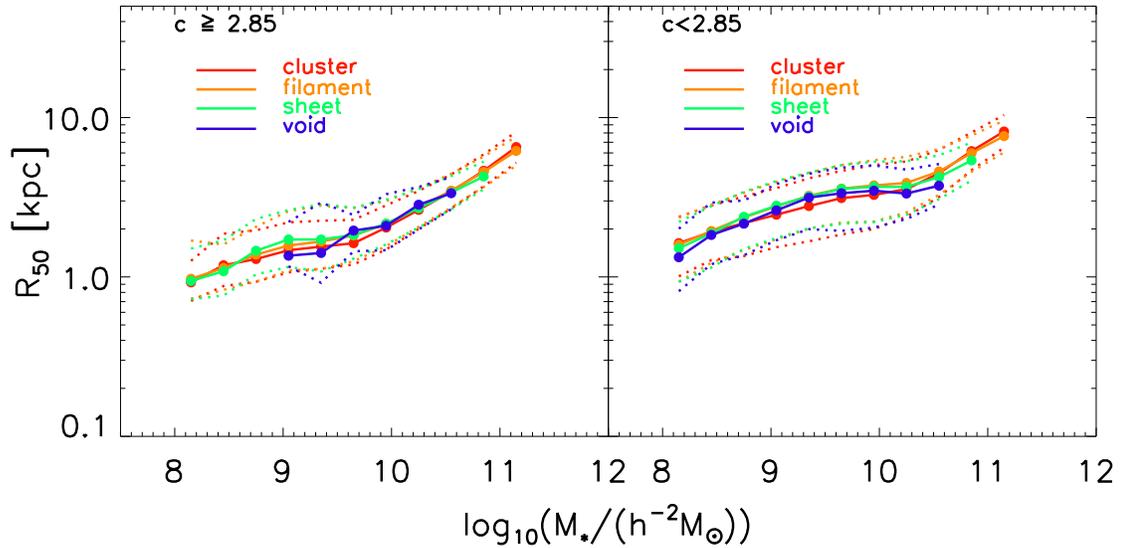}
\caption{ Mass-size relations of early- ($c\ge 2.85$) and late-type ($c<2.85$) galaxies in different environments, indicated by different colors: \cluster (red), \filament (orange), \sheet (green) and \void (blue).}
\label{fig:cw1}
\end{figure*}

\begin{figure*}
\center
\includegraphics[width=0.9\textwidth]{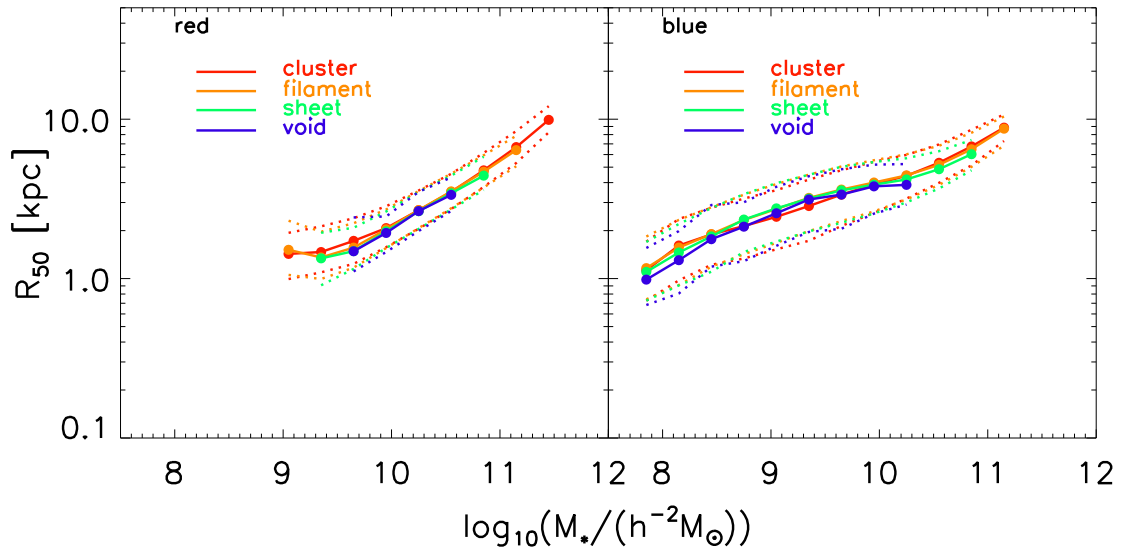}
\caption{ Same as Figure~\ref{fig:cw1}, but for red and blue galaxies.}
\label{fig:cw2}
\end{figure*}

\subsection{Dependence on cosmic web environment}

The second large scale environment we check is the cosmic web.  As
mentioned before, the galaxies in the NYU-VAGC samples can be
classified into one of four cosmic web types: \cluster, \filament,
\sheet, and \void, according to the positive number of the eigenvalues
of local tidal tensor $T_{i,j}$ constructed by \citet{Wang2012}. 

In this section, galaxies are further separated into early- and late-type 
galaxies according to their concentrations. Besides, galaxies are also 
divided into 'red' and 'blue' according to their $^{0.1}(g-r)$ colors: galaxies
with $^{0.1}(g-r)\geq 0.83$ are called red galaxies, while galaxies with 
$^{0.1}(g-r) < 0.83$ are called blue galaxies. The value $0.83$ roughly
corresponds to the bimodal scale in the color-magnitude relation.

Figure~\ref{fig:cw1} shows the mass-size relations of early- and late-type
galaxies, which are separated into different cosmic web types. The
solid points in different colors are the median sizes of galaxies in
different environments: \cluster~ (red), \filament~(orange), \sheet
~(green) and \void ~(blue).  As one can see, there is no
significant dependence of mass-size relations on the large-scale
environment either for early- or late-type galaxies. For early-type galaxies
in the mass range of $10^{9} \mstar \le M_{*} < 10^{10} \mstar$, the slight
difference of galaxies in different cosmic web types may be caused by the 
statistical uncertainty due to the sparse galaxies in each mass bin.
For late-type galaxies in the mass range of $10^{9} \mstar \le M_{*} < 10^{10}
\mstar$, we can barely see that the galaxies in the \cluster ~ are slightly smaller than those in other three cosmic web environments.

Figure~\ref{fig:cw2} shows the mass-size relations of red and blue galaxies
in different cosmic web environments. As one can see, there is almost no
difference for red or blue samples in different cosmic web types.

\subsection{Dependence on local number density}

\begin{figure*}
\center
\includegraphics[width=0.9\textwidth]{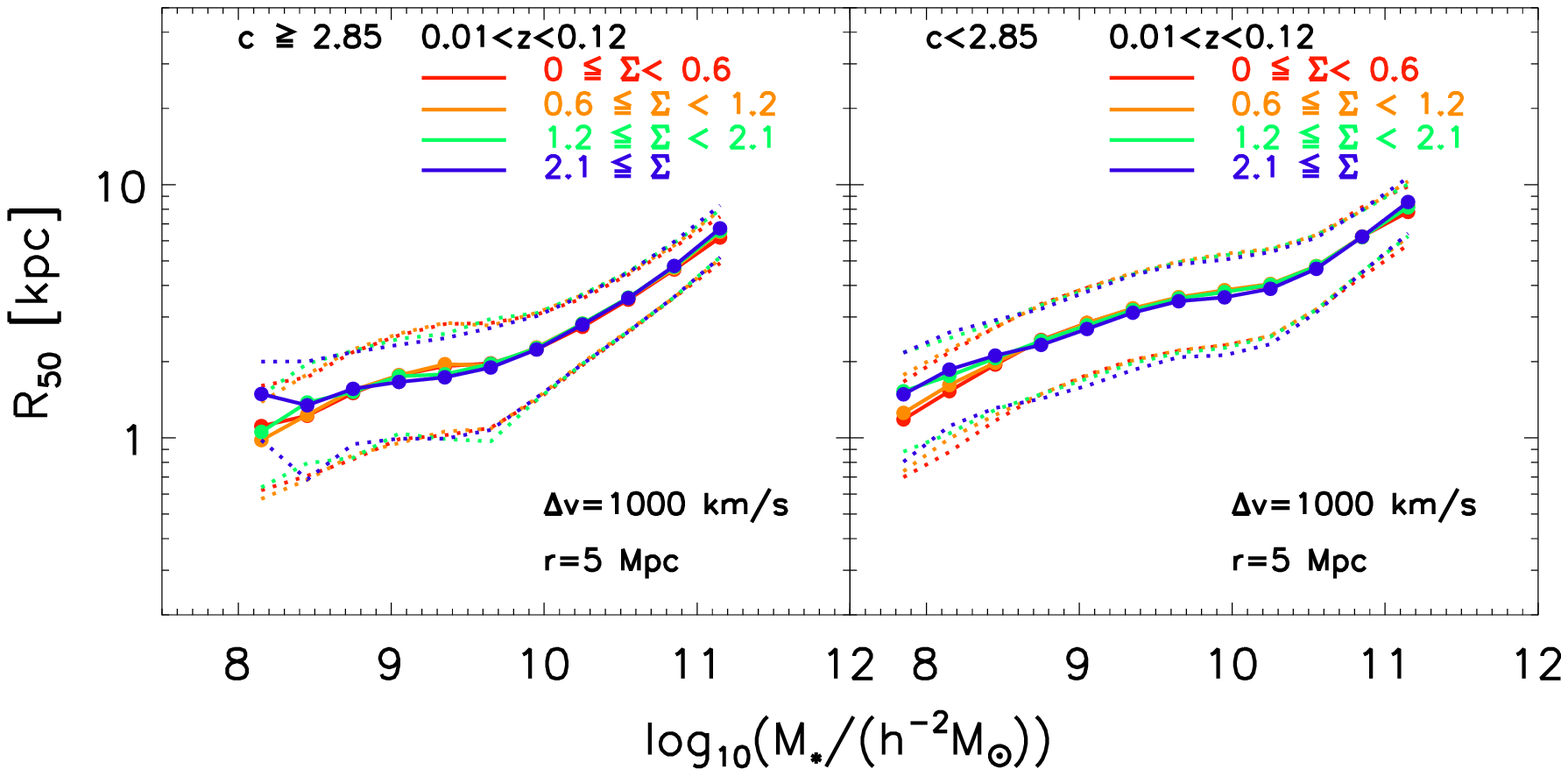}
\caption{Mass-size relations for early- and late-type galaxies 
  in different galaxy surface
  number density environments, measured by galaxy counts in cylinders
  with $r=5\,{\mpc}$ and $\Delta v = \pm 1000 \,{\rm km/s}$. }
\label{fig:density1}
\end{figure*}

\begin{figure*}
\center
\includegraphics[width=0.9\textwidth]{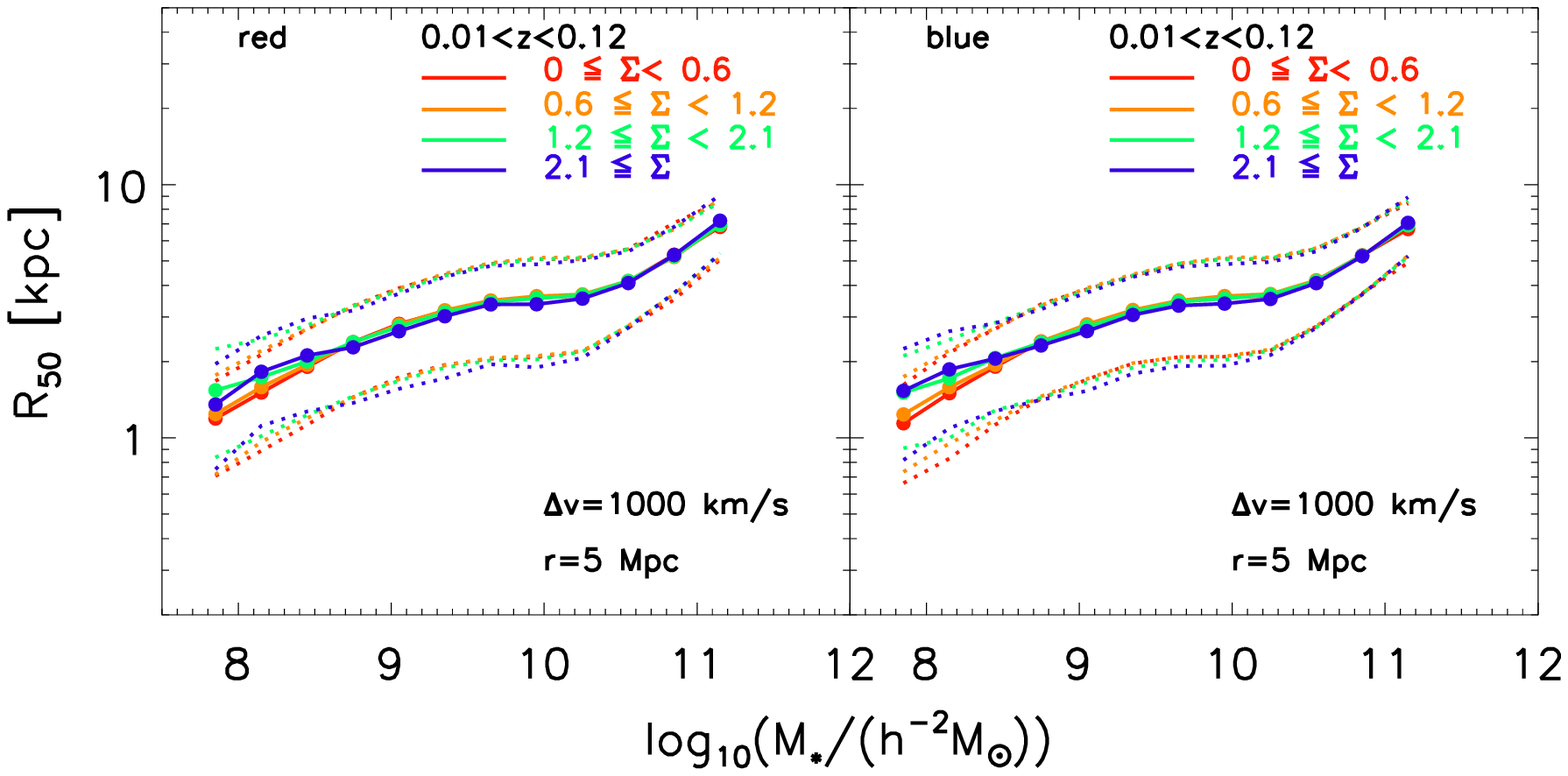}
\caption{ Same as Figure~\ref{fig:density1}, but for red and blue galaxies.}
\label{fig:density2}
\end{figure*}

The last large scale environment we check is the local galaxy number
density.  We investigate the dependence of the mass-size relation on
the local galaxy surface number density as defined in
Eq. \ref{eqn:density}.  We have used several sets of parameters of
$r=(1,2,5,10)\, {\mpc}$ and $\Delta v=\pm (500,1000) \,{\rm km/s}$ to
calculate the galaxy surface number density.  Our main conclusion is
not changed using different $r$ or $\Delta v$.  Based on the surface
number densities calculated using $r=5 \,{\mpc}$ and
$\Delta v = 1000 \,{\rm km/s}$, we divide galaxies into four equal
subsamples according to their surface number densities: (1)
$0<\Sigma<0.6$, (2) $0.6 \leq \Sigma<1.2$, (3) $1.2 \leq \Sigma<2.1$,
and (4) $\Sigma \geq 2.1$.  Besides, we have separated galaxies into
early- and late-type samples (red and blue samples) as been done
in Section 4.2. Figure~\ref{fig:density1} shows the mass-size relations 
of early- and late-type galaxies in different surface number density
environments.  We find that the sizes in different surface
number density environments are almost the same either for early-type
galaxies or for late-type galaxies. The slight difference for galaxies
with mass  $M_{*} < 10^{8.5} \mstar$ should be due to the statistical
uncertainty induced by the sparse number in these mass bins. Figure~\ref{fig:density2} shows the mass-size relations for red and blue galaxies.
As can be seen, the sizes of galaxies in different surface number density 
are also the same in the mass range $M_{*} \geq 10^{8.5} \mstar$.

\section{Summary}
\label{sec_summary}

Using a sample of $639,359$ galaxies selected from the SDSS DR7 and a
catalog of bulge-disk decompositions, we examine the size distribution
of galaxies and its dependence on the intrinsic properties of
galaxies, such as concentration, morphology, specific star formation
rate (sSFR), and bulge fraction, and on the
large-scale environments in the context of central/satellite
decomposition, halo environment, the cosmic web: \cluster, \filament,
\sheet ~and \void, as well as galaxy surface number density.

In order to investigate the dependence of the morphology, the galaxies
are separated into early- and late-type galaxies using the
concentration index $c=2.85$. Besides, we also separate galaxies into
elliptical and spiral galaxies using the morphological classifications
from GZ2. In addition to these, as the sSFR and bugle fraction of the
galaxy can be linked to the galaxy morphology, we also separate the
galaxies into low/high sSFR and high/low bugle fraction samples.

A double power law fitting formula has been used to quantify the
relations between the logarithm of the galaxy size and the $r$-band
absolute magnitude as well as the stellar mass. The related best
fitting parameters are provided in Table~\ref{table:parameters}. 
There is a clear trend that galaxy size increases with galaxy 
luminosity and stellar mass. Early-type (elliptical) galaxies have 
smaller sizes than late-type (spiral) galaxies. There is a strong 
dependence of luminosity (stellar mass)-size relation on the galaxy 
morphology, with steeper slopes for early-type (elliptical) galaxies. 
As expected, the size differences between low ($B/T<0.5$) and high
($B/T$) bulge fraction are somewhat similar to those of galaxies 
separated according to their morphological classifications. 
There is a trend of increasing slope and decreasing sizes for 
galaxies with high bulge fraction.

A number of efforts have been made to investigate the large scale
environmental dependence of the size distribution of galaxies. On one
hand, some studies claimed that there is no environmental dependence
in the mass-size relation \citep{Maltby2010, Nair2010, Huertas2013,
  Kelkar2015, Sara2017}. On the other hand, there are some studies
suggesting that the sizes of galaxies are dependent on their
environments\citep{Papo2012, Bass2013, Lani2013, Strazz2013,
  Delaye2014, Yoon2017}.

In this paper, we examine the environmental dependence of the
mass-size (luminosity-size) relations of galaxies in the SDSS DR7 with
much larger volumes and number of galaxies. Galaxies are separate into
centrals and satellites, and separated into high mass and low mass
halo environments. We do find galaxies in the stellar mass range
$10^{9.0} \mstar \le M_{*} < 10^{10.5} \mstar$ have a weak but
prominent halo environment dependence. The satellites and those in
massive halos have somewhat smaller sizes than their
counterparts. Beyond this stellar mass range, we do not see any halo
environmental dependence.

For the cosmic web, we find that there is almost no significant
difference in size for galaxies that are separated into four cosmic
web environments: \cluster, \filament, \sheet ~and \void.
Furthermore, we investigate the dependence of the mass-size relation
on the local galaxy surface number density.  Galaxies are then
separated into four equal subsamples according to their surface number
densities. We find that, galaxies in different surface number
densities have almost the same sizes.

\section*{Acknowledgements}

We would like to thank an anonymous referee for invaluable comments.
This work is supported by the 973 Program (No. 2015CB857002), and the
national science foundation of China (grant Nos. 11233005, 11621303).

This work is also supported by the High Performance Computing Resource
in the Core Facility for Advanced Research Computing at Shanghai
Astronomical Observatory.

Funding for the Sloan Digital Sky Survey IV has been provided by the
Alfred P. Sloan Foundation, the U.S. Department of Energy Office of
Science, and the Participating Institutions. SDSS acknowledges support
and resources from the Center for High-Performance Computing at the
University of Utah. The SDSS web site is www.sdss.org.

SDSS is managed by the Astrophysical Research Consortium for the
Participating Institutions of the SDSS Collaboration including the
Brazilian Participation Group, the Carnegie Institution for Science,
Carnegie Mellon University, the Chilean Participation Group, the
French Participation Group, Harvard-Smithsonian Center for
Astrophysics, Instituto de Astrof{\'i}sica de Canarias, The Johns
Hopkins University, Kavli Institute for the Physics and Mathematics of
the Universe (IPMU) / University of Tokyo, Lawrence Berkeley National
Laboratory, Leibniz Institut f{\"u}r Astrophysik Potsdam (AIP),
Max-Planck-Institut f{\"u}r Astronomie (MPIA Heidelberg),
Max-Planck-Institut f{\"u}r Astrophysik (MPA Garching),
Max-Planck-Institut f{\"u}r Extraterrestrische Physik (MPE), National
Astronomical Observatories of China, New Mexico State University, New
York University, University of Notre Dame, Observat{\'o}rio Nacional /
MCTI, The Ohio State University, Pennsylvania State University,
Shanghai Astronomical Observatory, United Kingdom Participation Group,
Universidad Nacional Aut{\'o}noma de M{\'e}xico, University of
Arizona, University of Colorado Boulder, University of Oxford,
University of Portsmouth, University of Utah, University of Virginia,
University of Washington, University of Wisconsin, Vanderbilt
University, and Yale University.

\label{lastpage}
\end{document}